\documentclass{aa}  

\usepackage{graphicx}
\usepackage{soul}
\usepackage{txfonts}

\newcommand{\oiii}{[O\,\textsc{iii}]}
\newcommand{\nii}{[N\,\textsc{ii}]}
\newcommand{\sii}{[S\,\textsc{ii}]}

\newcommand{\siii}{[S\,\textsc{iii}]}
\newcommand{\oii}{[O\,\textsc{ii}]}

\newcommand{\hii}{H\,\textsc{ii}}
\newcommand{\ha}{H$\alpha$}
\newcommand{\hb}{H$\beta$}
\newcommand{\cii}{[C\,\textsc{ii}]}

\newcommand{\revone}{}
\newcommand{\revtwo}{} 

\begin{document}

   \title{SDSS-V Local Volume Mapper (LVM): A Glimpse into Orion}


\newcommand{\OSU}{\label{OSU} Department of Astronomy, The Ohio State University, 140 West 18th Avenue, Columbus, Ohio 43210, USA}

\newcommand{\Alberta}{\label{Alberta} Department of Physics, University of Alberta, Edmonton, AB T6G 2E1, Canada}

\newcommand{\ANU}{\label{ANU} Research School of Astronomy and Astrophysics, Australian National University, Canberra, ACT 2611, Australia}

\newcommand{\IPAC}{\label{IPAC} Caltech-IPAC, 1200 E. California Blvd. Pasadena, CA 91125, USA}

\newcommand{\Carnegie}{\label{Carnegie} Observatories of the Carnegie Institution for Science, 813 Santa Barbara Street, Pasadena, CA 91101, USA}

\newcommand{\CCAPP}{\label{CCAPP} Center for Cosmology and Astroparticle Physics, 191 West Woodruff Avenue, Columbus, OH 43210, USA}

\newcommand{\CfA}{\label{CfA}Harvard-Smithsonian Center for Astrophysics, 60 Garden Street, Cambridge, MA 02138, USA}

\newcommand{\CITEVA}{\label{CITEVA} Centro de Astronomía (CITEVA), Universidad de Antofagasta, Avenida Angamos 601, Antofagasta, Chile}

\newcommand{\CNRS}{\label{CNRS} CNRS, IRAP, 9 Av. du Colonel Roche, BP 44346, F-31028 Toulouse cedex 4, France}

\newcommand{\ESO}{\label{ESO} European Southern Observatory, Karl-Schwarzschild Stra{\ss}e 2, D-85748 Garching bei M\"{u}nchen, Germany}

\newcommand{\ESOChile}{\label{ESOChile} European Southern Observatory, Avenida Alonso de Cordoba 3107, Casilla 19, Santiago 19001, Chile}

\newcommand{\HD}{\label{HD} Astronomisches Rechen-Institut, Zentrum f\"{u}r Astronomie der Universit\"{a}t Heidelberg, M\"{o}nchhofstra\ss e 12-14, D-69120 Heidelberg, Germany}

\newcommand{\ICRAR}{\label{ICRAR} International Centre for Radio Astronomy Research, University of Western Australia, 35 Stirling Highway, Crawley, WA 6009, Australia}

\newcommand{\IRAM}{\label{IRAM} Institut de Radioastronomie Millim\'{e}trique (IRAM), 300 Rue de la Piscine, F-38406 Saint Martin d'H\`{e}res, France}

\newcommand{\ITA}{\label{ITA} Universit\"{a}t Heidelberg, Zentrum f\"{u}r Astronomie, Institut f\"{u}r Theoretische Astrophysik, Albert-Ueberle-Str 2, D-69120 Heidelberg, Germany}

\newcommand{\IWR}{\label{IWR} Universit\"{a}t Heidelberg, Interdisziplin\"{a}res Zentrum f\"{u}r Wissenschaftliches Rechnen, Im Neuenheimer Feld 205, D-69120 Heidelberg, Germany}

\newcommand{\JHU}{\label{JHU} Department of Physics and Astronomy, The Johns Hopkins University, Baltimore, MD 21218, USA}

\newcommand{\Leiden}{\label{Leiden} Leiden Observatory, Leiden University, P.O. Box 9513, 2300 RA Leiden, The Netherlands}

\newcommand{\Maryland}{\label{Maryland} Department of Astronomy, University of Maryland, College Park, MD 20742, USA}

\newcommand{\MPE}{\label{MPE} Max-Planck-Institut f\"{u}r extraterrestrische Physik, Giessenbachstra{\ss}e 1, D-85748 Garching, Germany}

\newcommand{\MPIA}{\label{MPIA} Max-Planck-Institut f\"{u}r Astronomie, K\"{o}nigstuhl 17, D-69117, Heidelberg, Germany}

\newcommand{\Nagoya}{\label{Nagoya} Department of Physics, Nagoya University, Furo-cho, Chikusa-ku, Nagoya, Aichi 464-8602, Japan}

\newcommand{\NRAO}{\label{NRAO} National Radio Astronomy Observatory, 520 Edgemont Road, Charlottesville, VA 22903-2475, USA}

\newcommand{\OAN}{\label{OAN} Observatorio Astron\'{o}mico Nacional (IGN), C/Alfonso XII, 3, E-28014 Madrid, Spain}

\newcommand{\ObsParis}{\label{ObsParis} Sorbonne Universit\'{e}, Observatoire de Paris, Universit\'{e} PSL, CNRS, LERMA, F-75014, Paris, France}

\newcommand{\Princeton}{\label{Princeton} Department of Astrophysical Sciences, Princeton University, Princeton, NJ 08544 USA}

\newcommand{\UToledo}{\label{UToledo} University of Toledo, 2801 W. Bancroft St., Mail Stop 111, Toledo, OH, 43606}

\newcommand{\Toulouse}{\label{Toulouse} Universit\'{e} de Toulouse, UPS-OMP, IRAP, F-31028 Toulouse cedex 4, France}

\newcommand{\UBonn}{\label{UBonn} Argelander-Institut f\"ur Astronomie, Universit\"at Bonn, Auf dem H\"ugel 71, 53121 Bonn, Germany}

\newcommand{\UChile}{\label{UChile} Departamento de Astronom\'{i}a, Universidad de Chile, Camino del Observatorio 1515, Las Condes, Santiago, Chile}

\newcommand{\UConn}{\label{UConn} Department of Physics, University of Connecticut, Storrs, CT, 06269, USA}

\newcommand{\UCSD}{\label{UCSD}Center for Astrophysics and Space Sciences, Department of Physics,  University of California, San Diego, 9500 Gilman Drive, La Jolla, CA 92093, USA}

\newcommand{\UGent}{\label{UGent} Sterrenkundig Observatorium, Universiteit Gent, Krijgslaan 281 S9, B-9000 Gent, Belgium}

\newcommand{\ULyon}{\label{ULyon} Univ Lyon, Univ Lyon 1, ENS de Lyon, CNRS, Centre de Recherche Astrophysique de Lyon UMR5574,\\ F-69230 Saint-Genis-Laval, France}

\newcommand{\UMass}{\label{UMass} University of Massachusetts—Amherst, 710 N. Pleasant Street, Amherst, MA 01003, USA}

\newcommand{\UWyoming}{\label{UWyoming} Department of Physics and Astronomy, University of Wyoming, Laramie, WY 82071, USA}

\newcommand{\LAM}{\label{LAM} Aix Marseille Univ, CNRS, CNES, LAM (Laboratoire d’Astrophysique de Marseille), Marseille, France}

\newcommand{\UHawaii}{\label{UHawaii} Institute for Astronomy, University of Hawaii, 2680 Woodlawn Drive, Honolulu, HI 96822, USA}

\newcommand{\UCM}{\label{UCM} Departamento de F\'{\i}sica de la Tierra y Astrof\'{\i}sica, Universidad Complutense de Madrid, E-28040, Spain}

\newcommand{\IPARC}{\label{IPARC} Instituto de F\'{\i}sica de Part\'{\i}culas y del Cosmos IPARCOS, Facultad de Ciencias F\'{\i}sicas, Universidad Complutense de Madrid, E-28040, Spain}

\newcommand{\STScI}{\label{STScI} Space Telescope Science Institute, 3700 San Martin Drive, Baltimore, MD 21218, USA}

\newcommand{\McMaster}{\label{McMaster} Department of Physics and Astronomy, McMaster University, 1280 Main Street West, Hamilton, ON L8S 4M1, Canada}

\newcommand{\INAF}{\label{INAF} INAF -- Osservatorio Astrofisico di Arcetri, Largo E. Fermi 5, I-50157, Firenze, Italy}

\newcommand{\Sydney}{\label{Sydney} Sydney Institute for Astronomy, School of Physics A28, The University of Sydney, NSW 2006, Australia}

\newcommand{\CITA}{\label{CITA} Canadian Institute for Theoretical Astrophysics (CITA), University of Toronto, 60 St George St, Toronto, ON M5S 3H8, Canada}

\newcommand{\ASIAA}{\label{ASIAA} Institute of Astronomy and Astrophysics, Academia Sinica, No. 1, Sec. 4, Roosevelt Road, Taipei 10617, Taiwan}

\newcommand{\TKU}{\label{TKU} Department of Physics, Tamkang University, No.151, Yingzhuan Rd., Tamsui Dist., New Taipei City 251301, Taiwan}

\newcommand{\PSMA}{\label{PSMA} Penn State Mont Alto, 1 Campus Drive, Mont Alto, PA  17237, USA}

\newcommand{\ILL}{\label{ILL} ILL}

\newcommand{\stromlo}{\label{stromlo} Research School of Astronomy and Astrophysics, Australian National University, Mt Stromlo Observatory, Weston Creek, ACT 2611, Australia}

\newcommand{\UCatolica}{\label{UCatolica} Instituto de Astronom\'ia, Universidad Cat\'olica del Norte, Av. Angamos 0610, Antofagasta, Chile}

\newcommand{\UT}{\label{UT} McDonald Observatory, The University of Texas at Austin, 1 University Station, Austin, TX 78712-0259, USA}

\newcommand{\Vanderbilt}{\label{Vanderbilt} Department of Physics and Astronomy, Vanderbilt University, VU Station 1807, Nashville, TN 37235, USA}

\newcommand{\UNF}{\label{UNF} Department of Physics, University of North Florida, 1 UNF Dr. Jacksonville FL 32224}

\newcommand{\NAOC}{\label{NAOC} Chinese Academy of Sciences South America Center for Astronomy, National Astronomical Observatories, CAS, Beijing 100101, China}

\newcommand{\CASA}{\label{CASA} Center for Astrophysics and Space Astronomy, University of Colorado, 389 UCB, Boulder, CO 80309-0389, USA}

\newcommand{\UNAM}{\label{UNAM} Universidad Nacional Aut\'onoma de M\'exico, Instituto de Astronom\'ia, AP 106, Ensenada 22800, BC, M\'exico}

\newcommand{\UDP}{\label{UDP} Instituto de Estudios Astrof\'isicos, Facultad de Ingenier\'ia y Ciencias, Universidad Diego Portales, Av. Ej\'ercito Libertador 441, Santiago, Chile}

\newcommand{\Steward}{\label{Steward} Steward Observatory, University of Arizona, 933 N. Cherry Ave., Tucson, AZ 85721-0065, USA}  

\newcommand{\APO}{\label{APO} Apache Point Observatory and New Mexico State University, P.O.\ Box 59,
Sunspot, NM 88349-0059, USA}

\newcommand{\UNAMCU}{\label{UNAMCU} Universidad Nacional Aut\'onoma de M\'exico, Instituto de Astronom\'ia, AP 70-264, CDMX 04510, M\'exico}

\newcommand{\UWash}{\label{UWash}Department of Astronomy, University of Washington, Seattle, WA, 98195}

\newcommand{\CC}{\label{CC}Department of Physics, Colorado College, Colorado Springs, CO 80903}

\newcommand{\Utah}{\label{Utah}Department of Physics and Astronomy, University of Utah, 115 S. 1400 E., Salt Lake City, UT 84112, USA}

\newcommand{\UConcepcion}{\label{UConcepcion}Departamento de Astronom\'ia, Universidad de Concepci\'on, Casilla 160-C, Concepci\'on, Chile}

\newcommand{\FCLA}{\label{FCLA}Franco-Chilean Laboratory for Astronomy, IRL 3386, CNRS and Universidad de Chile, Santiago, Chile}

\newcommand{\Oklahoma}{\label{Oklahoma}Homer L. Dodge Department of Physics and Astronomy, University of Oklahoma, Norman, OK 73019, USA}

\newcommand{\UIUC}{\label{UIUC}Department of Astronomy, University of Illinois, Urbana, IL 61801, USA}

\newcommand{\Harvard}{\label{Harvard}Harvard-Smithsonian Center for Astrophysics, Cambridge, MA 02138, USA}

\newcommand{\caltech}{\label{caltech}Department of Astronomy, California Institute of Technology, Pasadena, CA 91125, USA}


\author{
       K. Kreckel\inst{\ref{HD}} \thanks{\email{kathryn.kreckel@uni-heidelberg.de}}  \and
       O. V. Egorov\inst{\ref{HD}}  \and
       E. Egorova\inst{\ref{HD}}  \and
       G. A. Blanc\inst{\ref{Carnegie}, \ref{UChile}} \and 
       N. Drory\inst{\ref{UT}} \and 
       M. Kounkel\inst{\ref{UNF}} \and
       J. E. M\'endez-Delgado\inst{\ref{HD}} \and
       C. G. Rom\'an-Z\'u\~niga\inst{\ref{UNAM}} \and
       S. F. S\'anchez\inst{\ref{UNAM}} \and
       G. S. Stringfellow\inst{\ref{CASA}} \and 
       A. M. Stutz\inst{\ref{UConcepcion}, \ref{FCLA}} \and 
       E. Zari\inst{\ref{MPIA}} \and
       J. K. Barrera-Ballesteros\inst{\ref{UNAM}} \and 
       D. Bizyaev\inst{\ref{APO}} \and 
       J. R. Brownstein\inst{\ref{Utah}} \and
       E. Congiu\inst{\ref{ESOChile}} \and 
       J. G. Fern\'andez-Trincado\inst{\ref{UCatolica}}\and
       P. Garc\'{\i}a\inst{\ref{NAOC}, \ref{UCatolica}} 
       \and
       L. Hillenbrand\inst{\ref{caltech}} \and
       H. J. Ibarra-Medel\inst{\ref{UNAMCU}} \and 
       Y. Jin\inst{\ref{Harvard}} \and
       E. J. Johnston\inst{\ref{UDP}} \and
       A. M. Jones\inst{\ref{STScI}} \and 
       J. Serena Kim\inst{\ref{Steward}} \and 
       J. A.\ Kollmeier\inst{\ref{CITA}, \ref{Carnegie}} \and
       S. Kong\inst{\ref{Steward}} \and
       D. Krishnarao\inst{\ref{CC}} \and
       N. Kumari\inst{\ref{STScI}} \and
       J. Li\inst{\ref{HD}} \and
       K. Long\inst{\ref{STScI}} \and
       A. Mata-S\'{a}nchez\inst{\ref{UNAMCU}} \and 
       A. Mej\'{\i}a-Narv\'aez\inst{\ref{Carnegie}} \and 
       S. Anastasia Popa\inst{\ref{HD}} \and 
       H-W Rix\inst{\ref{MPIA}} \and
       N. Sattler\inst{\ref{HD}} \and 
       J. Serna\inst{\ref{UNAM}, \ref{Oklahoma}} \and 
       A. Singh\inst{\ref{UChile}}  \and
       J. R. S\'anchez-Gallego\inst{\ref{UWash}} \and
       A. Wofford\inst{\ref{UNAM}, \ref{UCSD}} \and
       T. Wong\inst{\ref{UIUC}} 
}

\institute{\tiny
\HD      \and  
 \Carnegie \and
 \UChile \and
 \UT \and 
 \UNF \and  
 \UNAM \and
 \CASA \and 
 \UConcepcion \and
 \FCLA \and 
 \MPIA \and 
 \APO \and 
 \Utah \and 
 \ESOChile \and
 \NAOC \and
 \UCatolica \and 
 \caltech \and 
 \UNAMCU \and
 \Harvard \and
 \UDP \and
 \STScI \and
 \Steward \and
 \CITA \and 
 \CC \and
 \Oklahoma \and
 \UWash \and
 \UCSD \and
 \UIUC
}

   \date{Submitted March 12, 2024; accepted XX}


  \abstract
   {The Orion Molecular Cloud complex, one of the nearest (D = 406~pc) and most extensively studied massive star-forming regions, is ideal for constraining the physics of stellar feedback, but its $\sim$12~deg diameter on the sky requires a dedicated approach to mapping ionized gas structures within and around the nebula.  }
   {The Sloan Digital Sky Survey (SDSS-V) Local Volume Mapper (LVM) is a new optical integral field unit (IFU) that will map the ionized gas within the Milky Way and Local Group galaxies, covering 4300~deg$^2$ of the sky with the new LVM Instrument. }
   {We showcase optical emission line maps from LVM covering 12~deg$^2$ inside of the Orion belt region, with 195,000 individual spectra combined to produce images at 0.07 pc (35.3\arcsec) resolution. This is the largest IFU map made (to date) of the Milky Way, and contains well-known nebulae (the Horsehead Nebula, Flame Nebula, IC~434, and IC~432),  as well as ionized interfaces with the neighboring dense Orion B molecular cloud. }
   {We resolve the ionization structure of each nebula, and map the increase in both the \sii/\ha\ and \nii/\ha\ line ratios at the outskirts of nebulae and along the ionization front with Orion B. \oiii\ line emission is only spatially resolved within the center of the Flame Nebula and IC~434, and our $\sim$0.1~pc scale line ratio diagrams show how variations in these diagnostics are lost as we move from the resolved to the integrated view of each nebula. We detect ionized gas emission associated with the dusty bow wave driven ahead of the star $\sigma$ Orionis, where the stellar wind interacts with the ambient interstellar medium. The Horsehead Nebula is seen as a dark occlusion of the bright surrounding photo-disassociation region. This small glimpse into Orion only hints at the rich science that will be enabled by the LVM.
   }
   {}
   \keywords{ISM: general --
                HII regions --
                Galaxy: local insterstellar matter --
                ISM: clouds
               }
   \maketitle
%

\section{Introduction}

The Orion Complex is one of the largest and best studied star-forming regions in the Solar neighborhood, hosting also one of the most nearby OB associations \citep[see ][ and references therein]{bally08, ODell:2011, bouyalves15}. Renowned for its striking ionized nebulae, dense molecular clouds and a rich population of young star clusters at distinct stages of early evolution \citep{megeath2016, stutz2016, furlan2016}, it provides an unparalleled laboratory for studying the physics of star formation. Given its proximity (406~pc; \citealt{Menten2007, Kounkel2018,Binder:2018}), the Orion region has historically served as a test bed for theories of the sub-parsec scale interplay between gas, dust, and young massive stars \citep{Barnard1894, Sharpless1952, ODell1965, Becklin1967, Zuckerman1973, Tielens1985, Genzel1989,ODell:2003,ODell:2023}. 

Given its wide $\sim$12~deg diameter on the sky, modern studies have developed a comprehensive view across the entire region through rich multi-wavelength survey efforts.  
Molecular gas surveys have established a detailed view of multiple molecular lines across the bright cloud complexes \citep{Wilson2005, Kong2018, Stanke2022}. 
Dust mapping via stellar extinction was pioneered in the Orion region, tracing the 3D structures in the region \revone{\citep{Lombardi2011, Schlafly2015, Leike2020, Foley2023, Edenhofer2024}. }
Near to far-infrared surveys with space based instrumentation like the Spitzer, Herschel, and WISE observatories \citep{Megeath2012, Stutz2013, megeath2016}, and ground based facilities \citep[e.g.,][]{Hernandez2014, meingast16, Briceno2019, Grosschedl19} have built complete samples of young embedded protostars, tracking the earliest phases of star formation. \revone{Deep near-infrared and optical imaging surveys by Hubble have produced definitive studies of the stellar cluster, cataloging the stars, brown dwarfs, and planets \citep{Robberto2013, Robberto2020}.} 
With the advent of Gaia and other ongoing stellar spectroscopic surveys  (e.g., SDSS-IV APOGEE, \citealt{Majewski2017, Jonsson2020}) we have recently been able to develop rich datasets cataloging stars in the region, providing robust measurements of stellar distances, spectral types, and group kinematics \citep{stutz2016, fang2017, Kounkel2018, Cottle2018, Zari2019,Bailer-Jones:2021}. 

\begin{figure}
    \centering
    \includegraphics[width=3.5in]{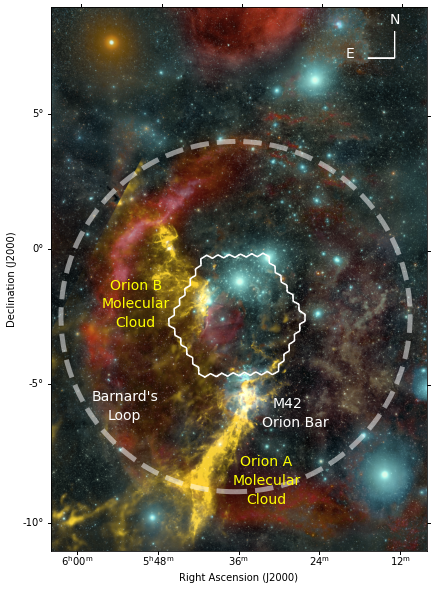}
    \caption{  
    Coverage of Orion. The glimpse into Orion provided by LVM in this work (solid white line) covers a large area in the Orion Constellation, including two of the three belt stars, but does not include M42. The ultimate goal of LVM is to achieve complete coverage (dashed white line) of the extended ionized bubble, with sparser coverage extending out to a radius of 19.5 degrees. Relevant large-scale features and regions are labeled for reference. The optical image is based on astrophotography, courtesy of Rogelio Bernal Andreo. Overplotted in yellow is the dust opacity map derived from Herschel and Planck data \citep{Lombardi2014}.} 
    \label{fig:constellation}
\end{figure}

Studies of the ionized gas in the region have been somewhat more limited, due to the challenge of carrying out continuum subtraction of bright (1st and 2nd magnitude) stars. Low angular resolution coverage of \ha\ emission was achieved with the $\sim$1~deg Wisconsin H-alpha Mapper (WHAM) Sky survey (WHAM; \citealt{Haffner1999, Haffner2003}), and a sharper $\sim$1~arcmin view obtained with the Southern H-alpha Sky Survey Atlas (SHASSA; \citealt{Gaustad2001}), with both surveys offering a general view of the large-scale features. The exception is the Orion Nebula (M42), including the Huygens region \citep{Huygens:1659} and ionizing Orion Nebula Cluster, that historically has been extensively studied \citep{ODell1965,Peimbert:1977,Baldwin:1991,Esteban:2004,Simon-Diaz:2006,Blagrave2007,MendezDelgado:2021b}. 
Optical integral field unit (IFU) spectroscopic maps would provide a natural link to the existing rich datasets, but existing instruments with wide field of view  (e.g., $\sim$1~arcmin with PPAK, VIRUS-P, MUSE) are inefficient when surveying such a large area of the sky. Small areas of Orion have been mapped by previous instruments \citep[e.g.,][]{Sanchez2007, Mesa-Delgado:2011,Weilbacher2015, McLeod2016, ODell2017, Fang2021}, but have generally focused on the bright areas of the Huygens region. 

As part of SDSS-V \citep{Kollmeier2017}, the LVM Instrument (LVM-I) has been designed as a survey telescope that can achieve a contiguous optical IFU survey of the Milky Way (MW) and Local Group targets \citep{drory24}. The instrument is designed to reach sub-pc scales in the MW, and 10~pc scales in the Magellanic Clouds. This effort to map the ionized gas and resolve individual Galactic nebulae has natural close ties to the ongoing efforts from SDSS-IV/APOGEE and SDSS-V/MWM to \revone{characterize} OB stars in the galaxy  \citep{2018ApJ...855...68R, 2019AJ....158...46B, 2020ApJ...891..107R, 2020ApJ...894....5R, 2020ApJS..247...17R, 2021ApJ...914...28M, Zari2021, 2023AJ....165...51R}. 
LVM will map Orion out to a radius of 6.5~deg (Figure \ref{fig:constellation}), starting from the geometric center of Barnard's Loop, north of M42 (which contains the Huygens region and is ionized by the Orion Nebula Cluster).

\begin{figure*}[t!]
    \centering
    \includegraphics[width=7in]{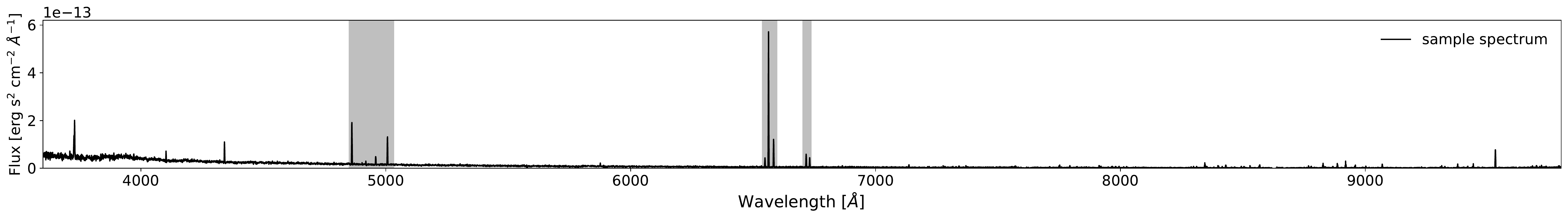}
    \includegraphics[width=3.5in]{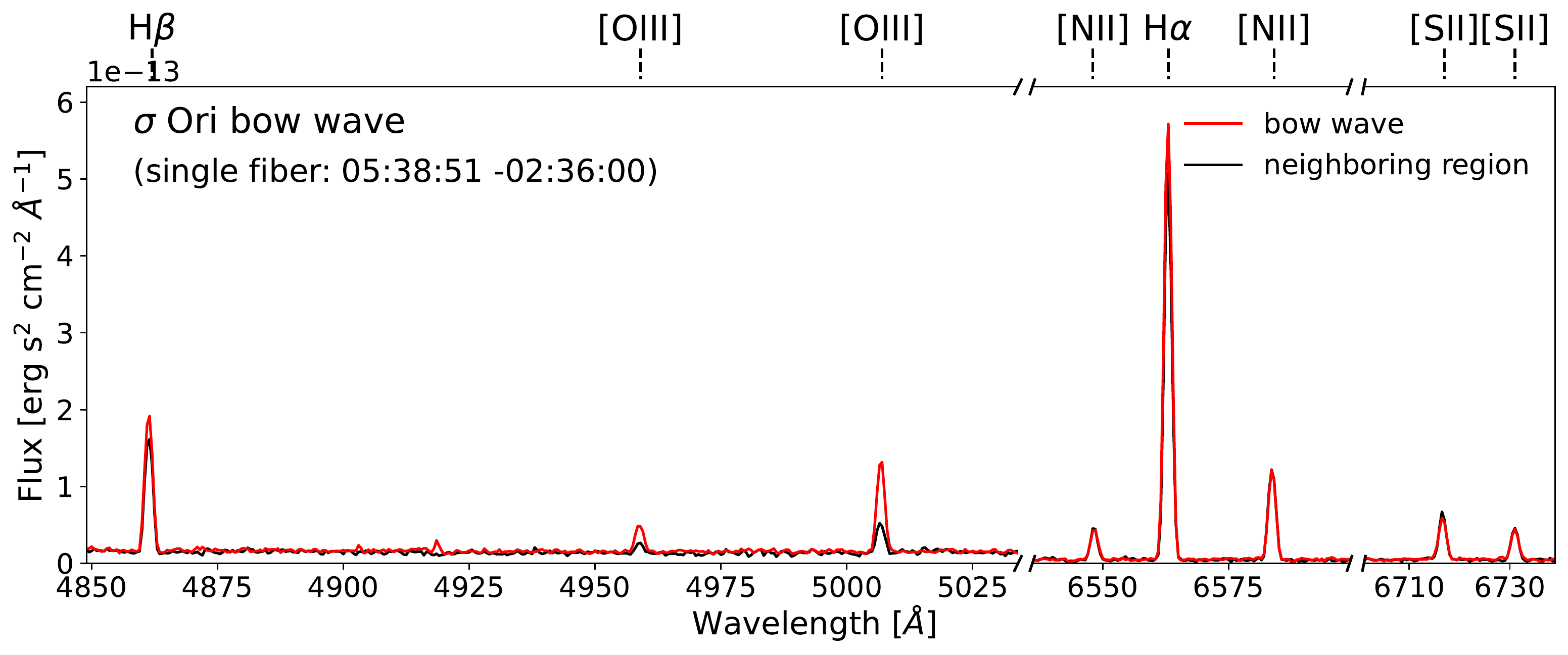}
    \includegraphics[width=3.5in]{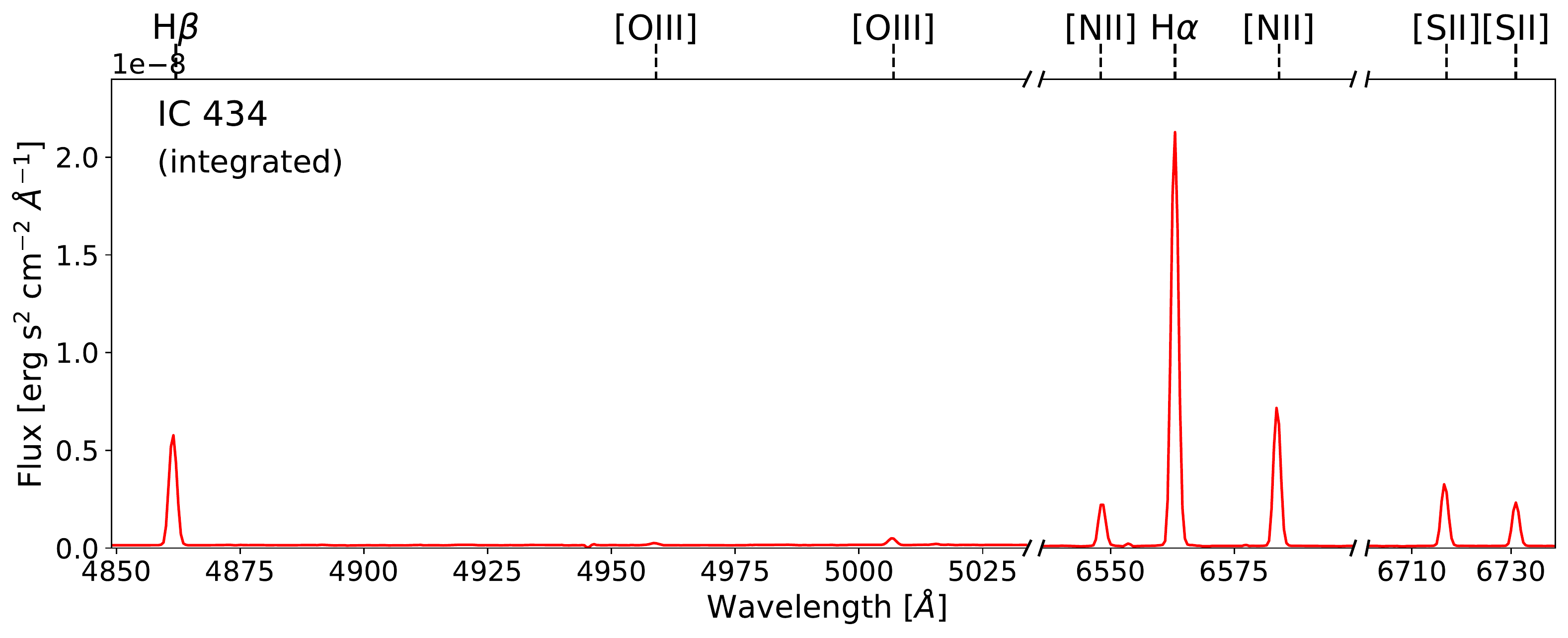}
    \includegraphics[width=3.5in]{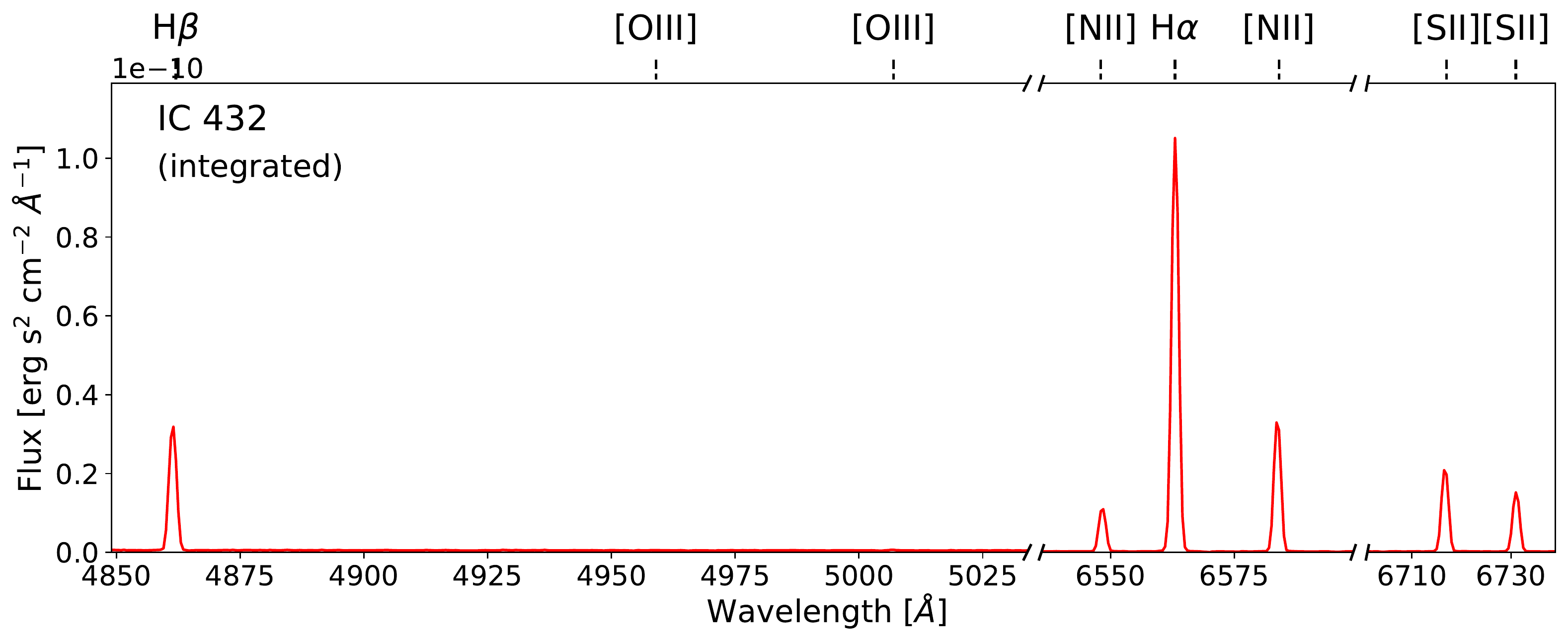}
    \includegraphics[width=3.5in]{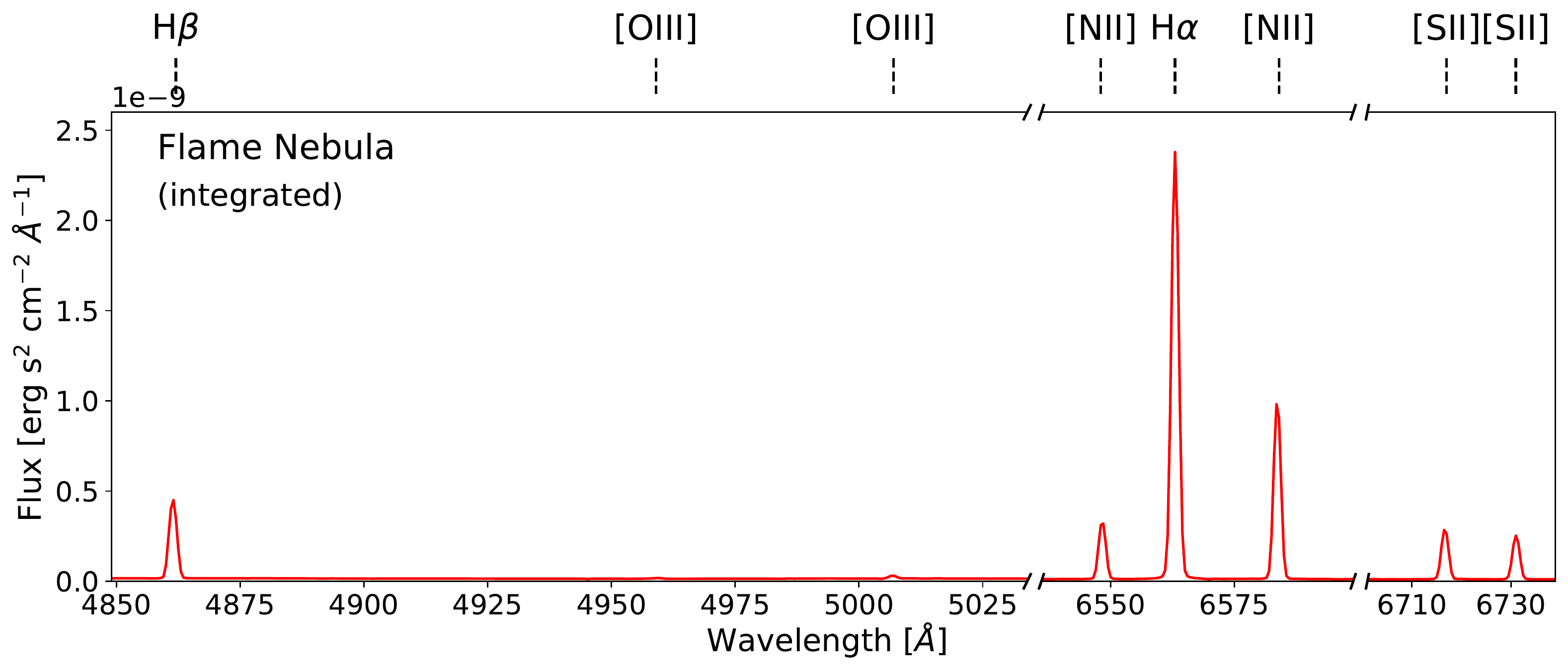}
    \caption{\revone{Sample LVM spectra. A full single fiber spectrum (top panel) demonstrates our full wavelength coverage from 3700-9800\AA, with the strongest emission line regions highlighted in grey. The lower panels highlight line emission from these narrow wavelength ranges, where H$\beta$, [OIII], [NII], H$\alpha$, and [SII] line emission is visible. A single fiber one the $\sigma$Ori bow wave (center left, red line) shows  no evidence of line broadening or multiple kinematic components, but clear enhanced Balmer and [OIII] emission compared to a neighboring spaxel (black), and an enhanced [OIII]/H$\beta$ line ratio compared to the integrated spectrum from IC~434. An integrated spectrum from IC~432 (bottom left) reveals no strong [OIII] emission but pronounced Balmer emission, consistent with photoionization by a B-type star. The Flame Nebula integrated spectrum (bottom right) shows nearly equal emission in both [SII] lines, indicating high electron densities.  }}
    \label{fig:spectra}
\end{figure*}

In this paper, we show a first glimpse into Orion, with our early survey tiles covering a  4.3~deg = 30~pc diameter region (assuming a distance of 406 pc; \citealt{Kounkel2018}) containing iconic Galactic nebular sources associated with the Orion B molecular cloud, such as the Horsehead Nebula and the Flame Nebula, in regions impacted by bright O-type stars including Alnitak ($\zeta$~Ori) and $\sigma$~Ori.  
We provide an overview of the data in Section \ref{sec:data}. We then explore the wealth of science enabled by the LVM in Milky Way star-forming regions by mapping the nebular conditions (Section \ref{sec:neb_conditions}), relating the ionizing sources to the ionized gas (Section \ref{sec:stars}), and charting the interfaces with dense molecular clouds (Section \ref{sec:interfaces}) using early LVM observations of the Orion region. At the end, we conclude in Section \ref{sec:conclusion}. Throughout this paper we use the following notation: \oiii\ to refer to \oiii$\lambda$5007, \nii\ to refer to \nii$\lambda$6583, \sii\ to refer to \sii$\lambda$6717+\sii$\lambda$6731, and \siii\ to refer to \siii$\lambda$9531.

\section{Data}
\label{sec:data}

   \begin{figure*}
   \centering
   \includegraphics[width=6in]{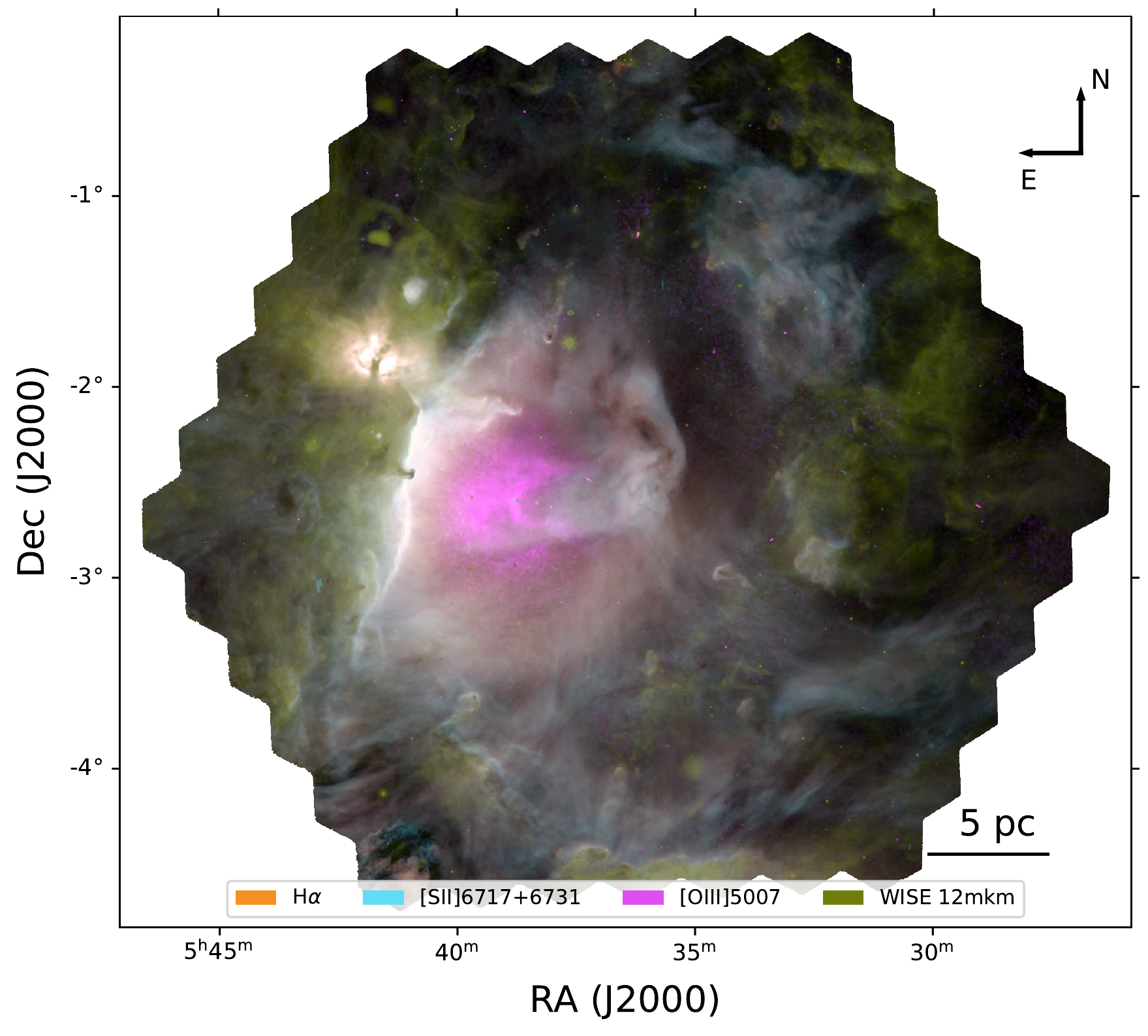}
   \caption{Our glimpse into Orion with LVM. 108 tiles cover 4~deg~$\times$~4~deg (30~pc $\times$ 30~pc), producing a total of almost 195,000 spectra. Colors show different emission lines (\ha: orange, \sii6717+6731: light blue, \oiii5007: magenta) trace the ionization structure of the nebula, and carve a bubble into surrounding dense gas (WISE 12$\mu$m: yellow). The ionized gas emission traces wispy, filamentary structures and dusty eroding clouds and clumps.  The iconic Horsehead Nebula sits at the bright filamentary ionization front and photo-disassociation region between the central ionized nebula IC~434 and the dark neighboring molecular cloud Orion B to the east, while  channels for escaping radiation are apparent in the north.   }
   \label{fig:orion_3color}%
    \end{figure*}
%


The LVM-I is a stable, wide field IFU telescope built and operated at Las Campanas Observatory (LCO\footnote{LCO is owned and operated by The Observatories of the Carnegie Institution for Science.}) by SDSS-V. It consists of four 16~cm telescopes, each equipped with bundles of fibers 35.3\arcsec diameter in size, which are fed to three DESI spectrographs (R$\sim$4000, 3700-9800\AA; \citealt{Perruchot2018}). The LVM-I was designed to efficiently survey the Milky Way and Local Group galaxies, and a detailed description of the science motivation and technical strategy are given in \cite{drory24}. The science data is obtained using a single telescope equipped with a 1801 fiber bundle, covering a $\sim$30 arcmin diameter hexagonal field of view. Through the use of a micro-lens-array, the total fill factor of these fibers is 83\%. The three additional telescopes enable simultaneous observation of spectrophotometric stars and sky fields. Data reduction is carried out via a pipeline that extracts and calibrates all spectra to produce wavelength and flux calibrated row-stacked spectra \citep{drory24}, removing the instrumental signatures from the data following the prescriptions described in \cite{Sanchez2006}. Our sky subtraction algorithm is still under development; it is carried out by modeling and extrapolating the sky from the sky field positions to the science field.  
However, given our focus in this work on bright lines bluewards of 7000$\AA$ the sky contamination is in most cases negligible.  \revone{Our proceedure of only carrying out Milky Way observations at high shadow height effectively limits the amount of geocoronal \ha\ emission present in our spectra \citep{drory24}. }

The final LVM survey of the Orion Nebula will eventually cover a 6.5~deg circular radius area on the sky, beyond which is planned a sparse grid of locations that have a 1/5th filling factor in order to map the Orion-Eridanus super-bubble out to a maximum radius of $19.5^{\circ}$. In this paper, we show results from the first 108 positions observed for the survey (`tiles'), covering a radius of $\sim$2~deg. These tiles were observed over 17 nights, comprising a total of almost 195,000 spectra. While dithers will be applied to recover emission falling in the gaps between fibers when observing some targets with the LVM (e.g., the Magellanic Clouds), no dithering is planned as part of the Milky Way survey, and so no dithering was performed for this set of observations \citep{drory24}. Assuming a distance of 406~pc \citep{Kounkel2018}, the 35.3\arcsec fibers yield spaxel diameters of 0.07~pc and total diameter for the field of view of $\sim$30~pc.

For each spectrum \revone{(Figure \ref{fig:spectra})}, we assumed a simple linear baseline and fit single Gaussians to a set of bright emission lines (\oii$\lambda$3727, \hb, \oiii$\lambda$5007, \ha, \nii$\lambda$6583, \sii$\lambda\lambda$6717, 6731, \siii$\lambda$9531), to obtain integrated line fluxes and line kinematics. This was sufficient given the low integrated stellar background contribution for a vast majority of the fibers. In addition, we also measured the flux by integrating the spectrum in a 6$\AA$ window centered on each line, subtracting the local continuum. While the resulting values are in very good agreement for these two methods, the maps obtained by simple integration were shallower and less affected by artifacts still present in the current version of the data, which can sometimes lead to imperfect line fits. Therefore, we use the integrated fluxes for all line ratio analysis in this paper, while the fitting results are used when we consider the ionized gas kinematics. We carried out image reconstruction following Shephard's method \citep{Sanchez2012}, a very simple and robust tessellation-free method to interpolate scattered data. A small number of individual dead fibers and fibers with low throughput were also masked, and interpolated over.  

Our Data Analysis Pipeline (currently under development) will handle this more precisely with a strategy designed for Resolved Stellar Populations (RSPs; \citealt{drory24}). As individual bright stars are contained within single 35.3\arcsec \ LVM fibers, including saturated 2nd and 4th magnitude stars $\zeta$~Ori and $\sigma$~Ori, they are not pronounced within our maps but can be seen as anomalous/masked values. With the LVM-I, we achieve a line-spread-function with $\sigma_{\rm ins}\simeq30$~km~s$^{-1}$ ($R\simeq$4000 at 6563~\AA), \revone{and for high signal-to-noise regions are able to centroid \ha\ to within 5~km~s$^{-1}$. However, due to ongoing developments in the data reduction pipeline, we do not yet obtain uniform measurements from pointing to pointing, making it challenging to produce kinematic maps covering the full field of view,} and as a result we present only limited results related to the emission line kinematics in this work.

The bulk of our analysis focuses on line ratios close in wavelength, such that reddening due to dust is negligible. When computing line ratios with wide separations (\siii/\sii) and extinction corrected \ha\ fluxes, we used the \texttt{pyneb} tool \citep{pyneb} to deredden all fluxes based on the observed \ha/\hb\ line ratio and assuming case~B recombination with \ha/\hb=2.86. We also assumed R$_V$ = 5.5 \citep{Mathis1981, Blagrave2007}, and a \cite{ccm} extinction law, and qualitatively our results do not change if we instead used R$_V$~=~3.1 as is commonly assumed for more diffuse regions \citep{Savage1979}. Although we adopted a fixed value for \ha/\hb, the theoretical value of this ratio is dependent on electron temperature and electron density \citep{Osterbrock2006}, and future work with LVM data will explore \revone{adapting} the case-B line ratio to the local conditions (as measured directly in the LVM data), along with using additional lines from the Balmer and Paschen series. 
These considerations do not significantly change the conclusions of this paper.   

Upon close inspection, artifacts and offsets are apparent (e.g small vertical stripes in the \nii/\ha\ maps, individual fibers with poor stellar continuum corrections in zoomed-in maps), particularly when comparing data between LVM neighboring pointings (e.g., in \ha/\hb\ maps). These are mainly due to current limitations in the data reduction and flux calibration, and ongoing work on the data reduction pipeline will improve these issues. As this paper mainly considered line ratios for lines close in wavelength, this has minimal impact on the results presented in this paper. When constructing line ratio maps, we minimized noise by applying a simple mask for all emission fainter than 1$\times$10$^{-17}$~erg~s$^{-1}$~cm$^{-2}$~arcsec$^{-2}$, approximately our 5$\sigma$ detection limit \citep{drory24}. 

Figure \ref{fig:orion_3color} shows our first glimpse into Orion with the LVM. 
It combines line emission from \ha\ (orange), \sii\ (blue), and \oiii\ (magenta), which fill the cavity that has been carved in the surrounding dense gas (traced by WISE 12$\mu$m emission in yellow; \citealt{Meisner2014, Roman-Zuniga2023}). 
The high ionization \oiii\ emission highlights individual photoionized bubbles within a sea of wispy, filamentary nebulosity. The iconic Horsehead nebula stands out against the vertical stripe of bright emission associated with the photo-disassociation region (PDR) at the interface to the neighboring molecular cloud Orion B (L1630; \citealt{Lynds1962}). Individual  dense cometary globules are also seen  dissolving as they are irradiated by the bright ionizing sources in these nebulae.  From the \ha\ morphology we see channels where radiation is able to escape and power the diffuse ionized gas (DIG), and it is clear that our symmetric modeling of \hii\ regions as simple spherical systems is not what is commonly found in nature \citep{Shields1990, Whitmore2011}.

\section{Nebular conditions}
\label{sec:neb_conditions}

   \begin{figure*}
   \centering
   \includegraphics[width=7in]{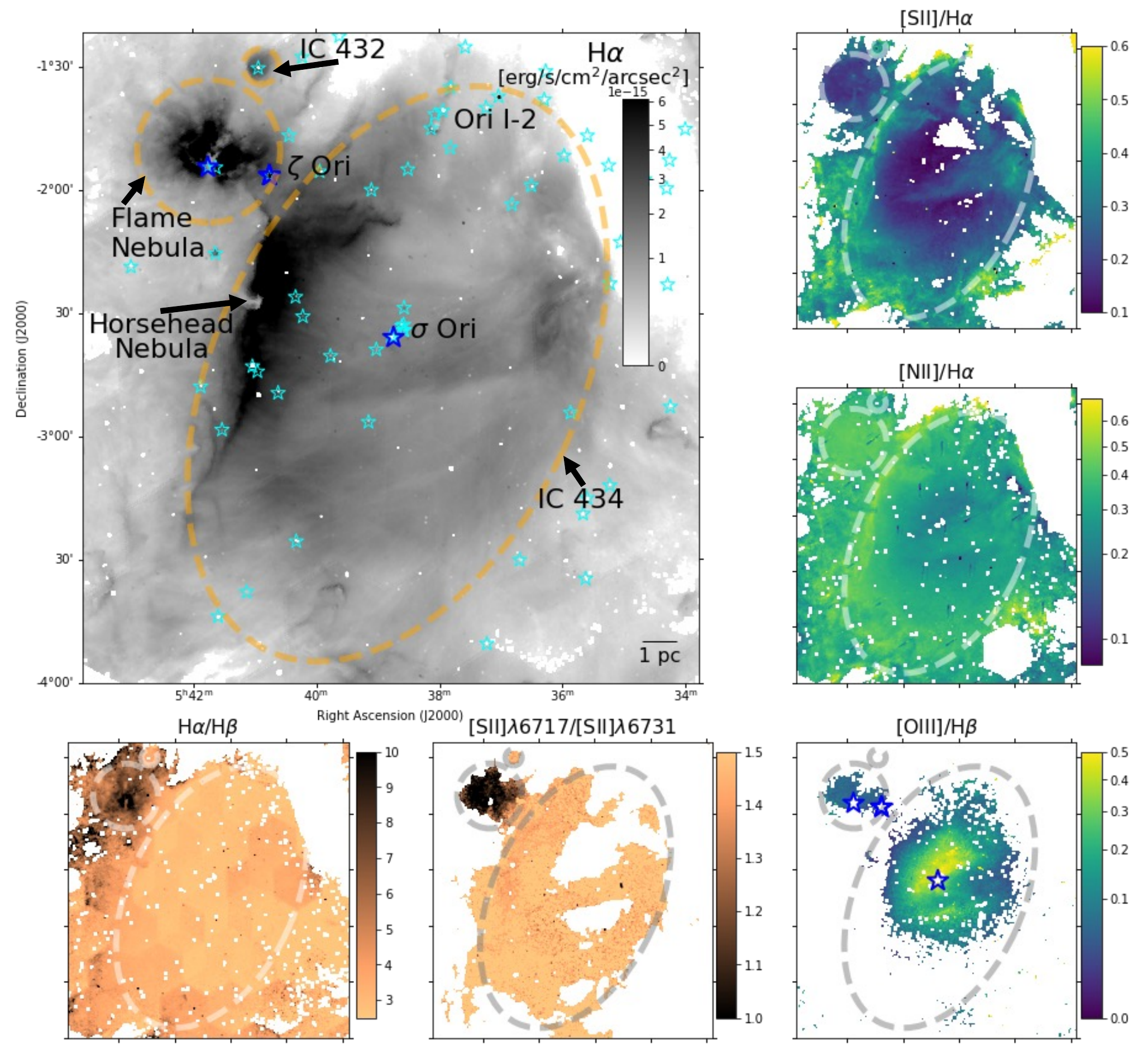}
   \caption{Line emission and line ratios mapped across 20~pc$\times$20~pc of the Orion belt region. Top left: \ha\ line emission is prominent within IC~434, the flame Nebula, and IC~432. 
   Star symbols mark all individual ionizing O-type (blue) and B-type (cyan) stars, \revone{and dashed lines show the approximate boundaries of each nebula}.  Missing individual pixels are masked fibers with bright stars where our simplistic stellar continuum fitting was insufficient.   Top center and right: \sii/\ha\  and \nii/\ha\ line ratio maps increase at the outskirts of IC~434 and the Flame nebulae. Bottom left: Dust extinction via the Balmer decrement increases along the extent of the neighboring Orion B molecular cloud, and peaks at the young embedded Flame Nebula. Bottom center: The increased dust concentration in the Flame Nebula is particularly pronounced in the \sii$\lambda$6717/\sii$\lambda$6731 line ratio, which traces changes in gas density.  Bottom right: \oiii/\hb\ line ratios can only be mapped in the centers of IC~434 and the Flame Nebula. Higher values towards the center of IC~434 reflect the changing ionizing structure of the nebula, with more highly ionized oxygen found closer to the ionizing source. }
              \label{fig:neb_conditions}%
    \end{figure*}

This IFU view into Orion provides access to multiple optical emission lines, which are commonly used to diagnose the physical conditions (temperature, density, chemical abundances) of the gas \citep{Pagel:1979,Baldwin1981, Vilchez88, Kewley2019}.  In this work, we focus on properties inferred from the brightest emission lines, specifically \hb, \oiii$\lambda$5007, \ha, \nii$\lambda$6583, \sii$\lambda\lambda$6717,6731, and \siii$\lambda$9531. All of these are sufficiently removed from atmospheric sky lines and telluric absorption to provide robust flux measurements even with our currently preliminary correction for these factors.  Fainter lines, such as auroral and recombination lines used to derive the electron temperature of the ionized gas, are also commonly detected in LVM targets and will be the focus of upcoming work.  

As is seen in Figure~\ref{fig:orion_3color}, emission lines from different ions emit brightly in different parts of the nebula (as labeled in Figure~\ref{fig:neb_conditions}), tracing the different ionization layers of the gas. In \ha\ emission (Figure \ref{fig:neb_conditions}, top left), three photoionized nebulae become apparent in relation to the ionizing O and B-type stars (queried from SIMBAD). The large IC~434, centered on the star $\sigma$~Ori, covers approximately 10~pc by 15~pc across the field of view. The bright Flame Nebula (NGC~2024) is more compact ($\sim$3~pc diameter), and marked by a pronounced vertical dust lane and centered on an embedded young stellar cluster \citep{Bik2003}. The smaller ($\sim$1~pc diameter) and fainter nebula IC~432 has been reported as a reflection nebula \citep{Magakian2003} (however see Section \ref{sec:IC432}), and is the only compact ionized region centered on a B-type star. 

\subsection{Mapping the ionization structure}

To explore in more detail the structure of these nebulae, we consider combinations of line ratios that diagnose changes in ionization conditions, and are close enough in wavelength that extinction corrections do not need to be applied.  The (\sii$\lambda$6717+\sii$\lambda$6731)/\ha\ 
(hereafter: \sii/\ha) and \nii$\lambda$6583/\ha\ (hereafter: \nii/\ha) line ratios (Figure \ref{fig:neb_conditions}, top center and top right) are fairly uniform across the Flame Nebula and IC~432, but show increasing values in outskirts of IC~434. This is indicative of a resolved ionization structure where the central cavity is filled with highly ionized gas, and a low-ionization outer shell that emits brightly in \nii\ and \sii. This radial structure is straightforward to model in one dimension by photoionization codes (e.g., CLOUDY; \citealt{Ferland2017}), however our two-dimensional view uncovers the elliptical morphology of this bubble. 
\revone{The eastern edge of IC~434 shows high \sii/\ha\ and \nii/\ha\ line ratios, and coincides with} the PDR separating IC~434 from the molecular cloud Orion B (see Section \ref{sec:interfaces}). Increased \sii/\ha\ and \nii/\ha\ structures are commonly found in regions dominated by the Warm Ionized Medium (WIM) in the Milky Way, and may be what we are detecting in the north and south-west edges of the map \citep{Haffner2009}.

   \begin{figure*}
   \centering
   \includegraphics[width=5in]{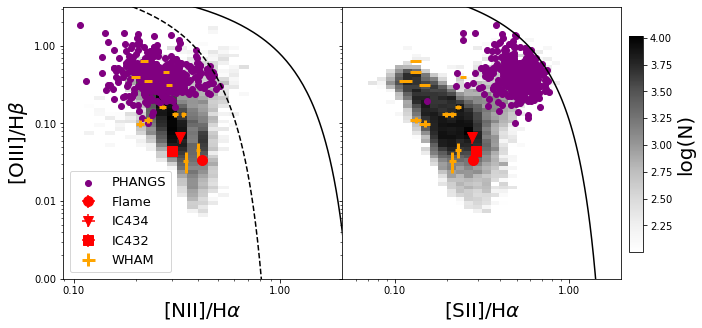}
   \caption{Diagnostic line ratios (left: \oiii/\hb\ vs \nii/\ha; right: \oiii/\hb\ vs \sii/\ha) for all pixels with significant detections in all lines. The curves shown from \cite{Kewley2001} (solid) and \cite{Kauffmann2003} (dashed) are commonly used (in integrated and kpc regions) to distinguish photoionized gas from gas ionized by harder ionizing sources.  Requiring the detection of \oiii\, this diagnostic is limited to the areas of the Flame Nebula and the center of NGC~434 (see Figure \ref{fig:neb_conditions}), as well as to regions consistent with photoionization. Integrated results for IC~434 (triangle), IC~432 (square), and the Flame Nebula (circle) are overplotted, as well as integrated measurements from WHAM for single O-star powered Milky Way \hii\ regions (orange) and extragalactic PHANGS \hii\ regions with matched \ha\ luminosity (purple). \revone{Error bars are included, but are typically smaller than the marker size.} }
              \label{fig:bpt}%
    \end{figure*}

The inner, high-ionization zone of IC~434 is detected in \oiii, and shows increased \oiii$\lambda$5007/\hb\ (hereafter: \oiii/\hb) line ratios towards the center of the nebula (Figure \ref{fig:neb_conditions}, bottom right). The more spherical appearance of \oiii/\hb, centered on $\sigma$~Ori, could indicate an increased homogeneity in the physical conditions at the interior of the nebula. High \oiii/\hb\ is also detected to each side of the dust-lane in the Flame nebula, where the dust extinction is less severe. 

This dust contribution can be directly mapped through two different line diagnostics. \revone{Under the assumption that the dust is located in the foreground along the line of sight,} then using the Balmer decrement (\ha/\hb; Figure \ref{fig:neb_conditions}, bottom right), we see high values at the location of the Flame Nebula and to the east, corresponding to the location of the molecular cloud Orion B. Assuming a \cite{ccm} reddening law and R$_V$~=~5.5 this corresponds to A$_V$~=~3-5~mag in this region.  
Using the \sii$\lambda$6717/\sii$\lambda$6731 density diagnostic (Figure \ref{fig:neb_conditions}, bottom center), we also see clear evidence for increased gas density  concentrated around the Flame Nebula. Here, low values of \sii$\lambda$6717/\sii$\lambda$6731 indicate high densities, with a value of 1 corresponding to densities of 10$^{3}$~cm$^{-3}$, and a value larger than 1.4 indicating low densities below 100~cm$^{-3}$ \citep{Draine2011}. In comparison, IC~434 and IC~432 show lower dust concentrations in both the Balmer decrement and the \sii\ density ratio, although distance estimates to the ionizing stars place them inside the Orion Molecular Cloud \citep{Schaefer2016, Bailer-Jones2021}, suggesting a lower concentration of dense gas and dust around these sites and less than 1~mag extinction in the V-band. 

These 0.1~pc scale tracers of the dust distribution provide a novel counterpart to the recent work mapping the 3D dust distribution in Orion \citep[e.g.,][]{Foley2023}. By placing synthetic ionized gas parcels at varying distances and velocities with respect to the existing 3D dust distribution models, it should be possible to compare mock observations with observed LVM line emission and line ratios. Variable levels of extinction should enable us to more precisely locate gas at both near and far distances along a single line of sight, and reconstruct the 3D ionized gas temperature and density structures, with the ultimate goal of refining our multi-phase reconstruction of the entire large-scale Orion region.  \revtwo{This reconstruction is quite challenging, with different geometric and kinematic orientations able to produce degenerate solutions, however extensive modeling and the use of many more emission lines beyond those from the Balmer decrement should provide a path towards unifying state of the art 3D dust maps with an understanding of 3D gas structures using the LVM. }  

\subsection{Comparing the resolved and integrated views}

   \begin{figure*}
   \centering
   \includegraphics[width=7in]{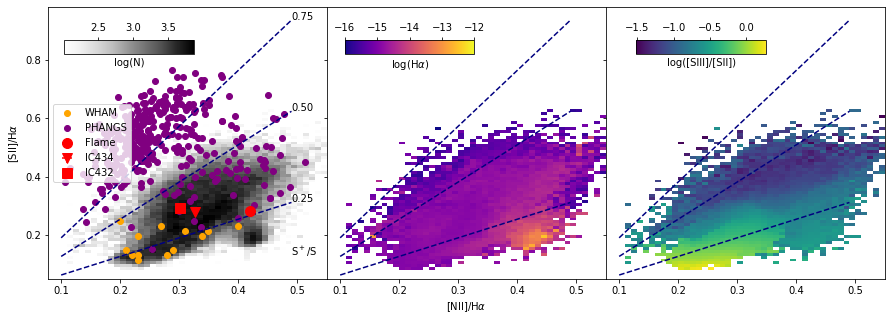}
\caption{ \sii/\ha\ as a function of \nii/\ha\  for pixels across the map (left). Considering the line emissivity models of \cite{Madsen2006} (navy lines), we see how line ratios are expected to change as a function of S$^{+}$/S (dashed).     We  color-code this 2D histogram distribution by the median extinction corrected \ha\ intensity (center) and the median extinction corrected \siii/\sii\ line ratio (right). We see that regions predicting lower S$^{+}$ abundances are well matched to regions with high S$^{++}$ abundances, as traced by \siii/\sii.  The Flame Nebula stands out, with high \nii/\ha\ but low \sii/\ha\ line ratios, and high \ha\ intensities. Integrated Milky Way \hii\ regions (IC~434: triangle, IC~432: square, Flame Nebula: circle, WHAM: orange points) exhibit low S$^{+}$/S values as much of the sulfur is ionized to S$^{++}$.  Extragalactic \hii\ regions at matched \ha\ luminosity (PHANGS: purple points) illustrate the challenges of separating \hii\ regions from the DIG on $\sim$100~pc scales, biasing diagnostic line ratios. }
              \label{fig:dig}%
    \end{figure*}

With the LVM, we will build up a sample of hundreds of \hii\ regions where we can explore correlations between resolved and unresolved line ratios, as well as their associations with individual ionizing sources. 
Combining multiple line ratios into diagnostic diagrams (e.g., BPT diagrams; \citealt{Baldwin1981}, \citealt{Veilleux1987}) can provide insight into the physical conditions of ionized gas, and are a common metric used for distinguishing photoionized gas from gas ionized by harder radiation fields (e.g., active galactic nuclei or shocks) in extragalactic studies. In both the \oiii/\hb\ vs \nii/\ha\ diagram (Figure \ref{fig:bpt}, left) and the \oiii/\hb\ vs \sii/\ha\ (Figure \ref{fig:bpt}, right), we plot the distribution of pixels in our field. 
Note that by requiring the detection of \oiii\ for these diagrams, we limit the coverage to the center of IC~434 and the Flame Nebula (see Figure \ref{fig:neb_conditions}), both of which appear consistent with pure photoionization from O and/or early B-type stars. The small number of points with slightly higher \sii/\ha\ correspond to the filamentary bar at the southern end of the full map (Figure \ref{fig:orion_3color}), which is $\sim$20~pc north of M42 and the Trapezium star cluster, and could reflect changes in the radiation field for gas closer to this very intensive ionizing source. 

To guide the eye, we also show the curve from \cite{Kewley2001}, based on photoionization models of integrated \hii\ regions, and the curve from \cite{Kauffmann2003}, based on empirical kpc scale measurements of galaxies. Physical interpretation of these lines in the context of resolved \hii\ regions is not straightforward \citep[for a more extensive discussion see][]{Sanchez2020}, \revone{and in this sub-pc resolved regime should mainly be used for context and to guide the eye}. 
\revone{However, these diagrams provide a novel counterpart to models. In a recent  
study focused on 3D fractal modeling single star \hii\ regions \citep{Jin2022}, the scattering of pixels on these diagrams arose from density variations within the nebula. However, as is also seen in our observations, they still found a correlated curved trend which follows the trend of the \cite{Kewley2001} line,  revealing that while density variations may increase scatter in this relation, to first order the underlying ionization parameter and metallicity still determine the line ratios.}  

Integrating over the entire nebulae (ellipses in Figure \ref{fig:neb_conditions}), we summed all spectra within the boundaries of IC~434, IC~432, and the Flame Nebula \revone{(Figure \ref{fig:spectra}) after masking the bright O and B stars (denoted as cyan and blue stars on Figure \ref{fig:neb_conditions})} and then performed Gaussian fits to the resulting strong emission lines. This approach is equivalent to weighting the integrated line flux by the intensity,  such that the integrated line ratios do not necessarily follow the peak of the distribution when looking at the resolved maps (red points, Figure \ref{fig:bpt}). The integrated \oiii/\hb\ values in IC~434 and the Flame Nebula do match reasonably well with the distribution, as that emission is centrally concentrated and peaks in the brightest regions. However, both \nii/\ha\ and \sii/\ha\ have somewhat higher values,  because most of the bright emission from \nii\ and \sii\ is located in the outer shell of the nebula. \revone{The compact IC~432, photoionized by a B-type star, is an outlier and sits at very low \oiii/\hb, as is expected given its softer ionizing spectrum.}   

We compare these regions with extragalactic \hii\ regions drawn from the PHANGS (Physics at High Angular resolution in Nearby GalaxieS) nebular catalog \citep{Groves2023}, selecting only those \hii\ regions that have similar \ha\ luminosities to what we measure for IC~434 and the Flame Nebula ($\sim 1 \times 10^{36}$~erg~s$^{-1}$). These are some of the faintest regions in the PHANGS catalog, making up less than 2\% of the catalog, and as a result these \hii\ regions have a strong selection bias against low \oiii/\hb\ values due to S/N limitations on \oiii.  Extragalactic \hii\ region searches do not routinely recover these individual O-star \hii\ regions, and in fact at the 100~pc scales achieved by the PHANGS, regions such as this would not easily be distinguished from the neighboring bright M42 region.  

Integrated Milky Way \hii\ regions were also observed as part of the  Wisconsin H-Alpha Mapper (WHAM) survey \citep{Haffner2003}. These regions, associated with individual O-type stars  \citep{Madsen2006}, are overplotted in orange, with all line ratios converted from photon flux ratios to energy flux ratios. As WHAM only reports the \sii$\lambda$6717 line, we assume \sii$\lambda$6731 = $\frac{1}{1.45} \times$\sii$\lambda$6717, which is true if the gas is in the low density limit ($n_e<100$ cm$^{-3}$). If the density is higher ($n_e\sim1000$ cm$^{-3}$), the line intensity ratio could be a factor of $\sim$2 higher \citep{Osterbrock2006}. The WHAM Milky Way \hii\ regions generally agree very well with the range of line ratios observed in our data, with all regions sitting at high \oiii/\hb\ line ratios associated with very early O-type stars (O4-O6), and hence harder ionizing radiation.

By focusing only on the \sii/\ha\ and \nii/\ha\ diagnostics, we are not limited by the faint \oiii\ line and can achieve a more comprehensive view of conditions across our field of view. We observe a general correlation between these two diagnostics (Figure \ref{fig:dig}), and compare with simple models derived in \cite{Madsen2006} for understanding the ionization sources powering diffuse ionized gas in the Galaxy. 
These models hold when you can assume that the N$^{+}$/N is constant (i.e., N$^{++}$ and N$^{0}$ are not dominant), which we expect to be the case in most of our map as \oiii\ is only detected in the center of IC~434 and the Flame Nebula.  
\cite{Madsen2006} found that the relative variations between  \sii/\ha\ and \nii/\ha\ can be explained by variations in the fraction of singly ionized sulphur (S$^{+}$/S). In our LVM map we find that values ranging from 20-75\% are well able to explain our data. 
\revone{We span a fairly narrow range in \nii/\ha, which in the \cite{Madsen2006} model would correspond to an electron temperature of 6000-8000~K (not included in Figure \ref{fig:dig}) and consistent with what is expected in Orion \citep{Wilson2015}. } 
Different sub-structures are visible in the distribution, and are associated with different nebular regions. Colorcoding the binned 2D histogram by the median extinction corrected \ha\ intensity (Figure \ref{fig:dig}, center), we see the points at high \nii/\ha\ and lower \sii/\ha\ correspond to the bright Flame Nebula, and are associated with fairly low S$^{+}$/S as most of the sulfur is likely present as S$^{++}$. This is consistent with a high abundance of O$^{++}$ in this young region, as is also mapped directly by the bright \oiii\ intensity (Figure \ref{fig:orion_3color}). 
The central cavity of IC~434 is also apparent in these diagrams, particularly when color-coding the binned 2D histogram by the median extinction corrected \siii/\sii\ line ratio (Figure \ref{fig:dig}, right). It stands out with high \siii/\sii, in good agreement with the modeled low S$^{+}$/S as there is clearly a high fraction of S$^{++}$ present, and is in contrast to the outer regions of the nebula where higher values of \nii/\ha, \sii/\ha, and lower values of \siii/\sii\ are seen. 
Integrated Milky Way \hii\ regions (IC~434: triangle, IC~432: square, Flame Nebula: circle, WHAM: orange points) exhibit significantly lower S$^{+}$/S values than extragalactic \hii\ regions with similar \ha\ luminosity (PHANGS: purple points).  This could reflect a systematic bias in these faint extragalactic regions, as on $\sim$100~pc scales it remains challenging to fully separate DIG emission from \hii\ regions, and could artificially elevate the \sii/\ha\ line ratios and hence appear as an overall higher S$^{+}$/S abundance.

\section{Ionized gas and their ionizing stars}
\label{sec:stars}

Compared to extragalactic studies, the two-dimensional resolved emission line maps obtained with the LVM provide the unique opportunity to directly link ionized nebulae to the individual stars powering individual bubbles.   
On the \ha\ image in Figure \ref{fig:neb_conditions}, we overlay all cataloged O and B-type stars. The three O-type stars present in the field are $\zeta$~Ori and $\sigma$~Ori --two of the brightest stars in the sky--, plus one heavily embedded O-type star (IRS2b, the fainter companion of IRS2) at the center of the Flame nebula that was only detected with infrared spectroscopy \citep{Bik2003}. IC~432 also has a clear correspondence with a single B-type star. Below, we discuss the relation between stars and gas for each of these objects.  

\subsection{$\sigma$ Orionis}

   \begin{figure*}
   \centering
   \includegraphics[width=7.5in]{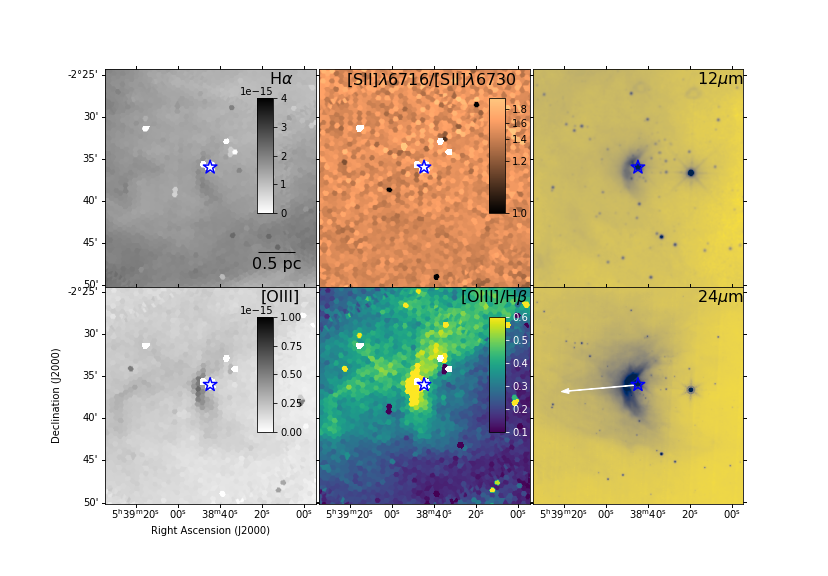}
   \caption{$\sigma$ Orionis (blue star), the central ionizing source of IC~434,  and its bow wave. The star system consists of a close binary of spectral type O9.5V and B0.5V with a third massive companion \citep{simon-diaz2011}, and is moving eastward through the nebula in agreement with the arc direction (white arrow, representing 0.1~Myr of travel time; \citealt{Perryman1997}). Apparent in imaging in \ha\ (top left), \oiii\ (bottom left), WISE 12$\mu$m (top right), and Herschel MIPS 24$\mu$m (bottom right), we see the wind and radiation pressure interacting with the surrounding ISM \citep{Ochsendorf2014}. We further report increased  \oiii/\hb\ emission (bottom center) but no increase in the gas density as traced by the \sii\ line ratio (top center).   }
              \label{fig:sigma_orionus}%
    \end{figure*}

Given the $\sim$10~pc diameter of IC~434 and the relatively weak stellar wind \citep{Najarro2011}, it is estimated that $\sigma$~Ori alone is unable to have driven the bubble we see \citep{Ochsendorf2015}. The system is kinematically associated with stars in the spatially and kinematically identified Orion C stellar complex \citep{Kounkel2018}, which contains sub-groups with median ages of 5-7~Myr. This, along with the high proper motions measured for $\sigma$~Ori \citep{Perryman1997}, have led to the suggestion that $\sigma$~Ori has moved over its lifetime from the bubble edge into the cavity, where it illuminates IC~434 and continues to approach the eastern Orion B molecular cloud, L1630 \citep{Ochsendorf2015}. 

This motion through the ionized cavity is evinced by the leading infrared-bright arc visible at $\sim$0.1~pc distance from the star in infrared wavelengths \citep{Caballero2008}. 
If a star is moving supersonically with respect to the local gas, a wave forms at the location where the wind and ambient ISM interact and sweep up material \citep{vanBuren1990, MacLow1991, Henney2019a}. \cite{Ochsendorf2014} carried out a detailed study of the infrared emission associated with this wave, modeling the dust grain properties associated with this dust structure and the physics of the wind-ISM interaction, demonstrating how such data can be used to constrain the dust grain size distribution and composition. 

We report here corresponding \ha\ and \oiii\ features associated with this  wave (Figure \ref{fig:sigma_orionus}), which exhibits high \oiii/\ha\ line ratios \revone{(Figure \ref{fig:spectra})}, and which could not be resolved in existing \ha\ imaging from the Southern H-Alpha Sky Survey Atlas (SHASSA; \citealt{Gaustad2001}). We see no corresponding change in the \sii\ intensity, or any change in the \sii\ density ratio that might suggest the presence of high ($>$100~cm$^{-3}$) density gas, at the location of the wave. \revone{We also see no evidence for broadened line emission or double-peaked spectral line shapes, typical expansion or outflow signatures (Figure \ref{fig:spectra}). }
The association of \ha\ and \oiii\ emission with this leading wave suggests that what we observe is a bow wave (containing both dust and gas), and not a gas-free dust wave, information that can be used for modeling the wind strength from this star \citep{Henney2019c}. The optical bow wave is in good agreement with the direction of motion of the star, as indicated by the white arrow.

\subsection{IC 432}
\label{sec:IC432}

\begin{figure}
   \centering
   \includegraphics[width=3.5in]{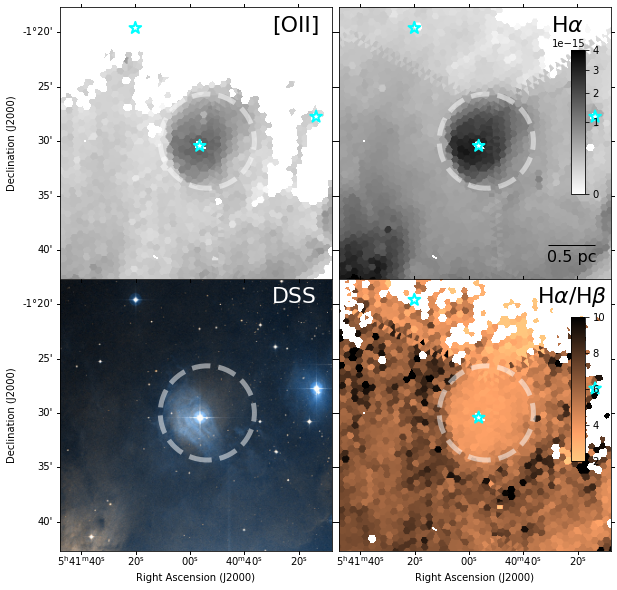}
   \caption{Line emission and optical images of IC~432. Cataloged as a reflection nebula, is revealed to also exhibit extended line emission in multiple lines, including \oii~ and \ha~ (top row). The bright, blue reflection nebula is also seen to the east in optical Digitized Sky Surveys (DSS) red and blue imaging (bottom left), however, there is relatively low extinction inferred via the Balmer decrement (\ha/\hb, bottom right). The central star, HD 37776 (cyan), is a magnetic chemically peculiar (helium-strong) star of spectral type B2V \citep{Thompson1985}.  }
              \label{fig:IC432}%
    \end{figure}

Classified in \cite{Magakian2003} and SIMBAD 
as a reflection nebula, where light from a nearby star is reflected by dust grains, we observe that IC~432  also exhibits strong, extended ($\sim$1~pc diameter) line emission (Figure \ref{fig:IC432}). This emission is slightly offset from the blue reflected light, and is well centered on HD~37776 (V901 Ori). 
This star is classified as a magnetic chemically peculiar (helium-strong) star of spectral type B2V \citep{Thompson1985, Mikulavek2008, Kochukhov2011}. It is the only B-type star in our field that is clearly associated with its own ionized nebula, perhaps due to its proximity to the nearby dense Orion B molecular cloud L1630. However it appears to be distinct from the dense cloud, as is apparent by its relatively low optical extinction in the Balmer decrement maps (Figure \ref{fig:IC432}, bottom left), and exhibits significantly lower density as judged by the \sii\ density diagnostic (Figure \ref{fig:neb_conditions}, bottom center).   In BPT diagnostics (Figure \ref{fig:bpt}), its line ratios appear consistent with photoionization and lie close to the O~star dominated nebulae IC~434 and Flame Nebula.

\subsection{Flame Nebula (NGC~2024)}

The Flame Nebula hosts the youngest population of young stellar objects (YSOs) in the Orion cloud \citep{Meyer1996, Levine2006}, and is heavily embedded (A$_V \sim$ 30 mag; \citealt{Lenorzer2004}). \cite{vanTerwisga2020} identified two young populations embedded within the nebula by characterizing their protoplanetary disk systems, finding an eastern population with ages of $\sim$0.5~Myr and a western population with ages 1~Myr (see also \citealt{Stutz2013}). The central ionizing source, associated with the infrared source IRS2b (the fainter component of the famous IRS2), has been classified with K band spectroscopy as having a spectral type O8 V \citep{Bik2003, Kandori2007}. Authors have claimed that additional UV ionizing sources remain to be identified in this heavily obscured region, or that other sources identified outside of the core of the nebula (e.g., IRS 1; B0.5; \citealt{Burgh2012}) may also contribute significantly to the total ionizing flux. 
From our maps of the dust reddening and \sii\ density diagnostic (Figure \ref{fig:neb_conditions}), it is clear that this region is the most heavily embedded. With a relatively smooth \sii$\lambda$6717/\sii$\lambda$6731 line ratio of $\sim$1.0 this corresponds to gas densities of almost 1000~cm$^{-3}$ \citep{Osterbrock2006}, and supports the assumption that this is a ionization-bounded nebula \citep{Bik2003}.

\subsection {Alnitak ($\zeta$ Orionis)}

The third known O-type star in our field is Alnitak, a belt star associated with the 2nd magnitude $\zeta$ Orionis system. This is composed of a close binary containing an O supergiant star and a B-type star (spectral types O9.5 Iab and B0.5 IV), and a wide B-type companion \citep{Hummel2013}. 
$\zeta$ Ori is an interesting case as it has no obvious ionized gas associated with it. Located (in projection) between the Flame Nebula and IC~434, it also has a wide range of distances reported in the literature. From Hipparcos a parallax distance of 225$^{+38}_{-27}$~pc was reported \citep{vanLeeuwen2007}, placing it over 100~pc closer to us than the Orion~C group \citep{Kounkel2018}. Too bright for Gaia, \cite{Hummel2013} used spectroscopy of the lower mass stars in the triple system to infer a photometric distance of 387 $\pm$ 54~pc. Given its radial velocity information, this places the star in good agreement with the Orion~D group \citep{Kounkel2018}, which exhibits generally older ages $\sim$5~Myr. \cite{Hummel2013} also find an age of 7~Myr for the lower mass component of the system, so it is possible this star has already dispersed or migrated away from its natal cloud. 
Given the projected separation from the Orion B molecular cloud, it seems likely that ionizing radiation from this source is able to propagate widely into the surrounding region and contribute more broadly to the diffuse ionized emission observed in the map. $\delta$~Ori, another O~type star associated with the Orion~D complex and located $\sim$5~pc north of our map, has also been noted as having no clear \hii\ region associated \citep{Reynolds1979}.

\section{Interfaces with dense clouds}
\label{sec:interfaces}

   \begin{figure*}
   \centering
   \includegraphics[width=6.5in]{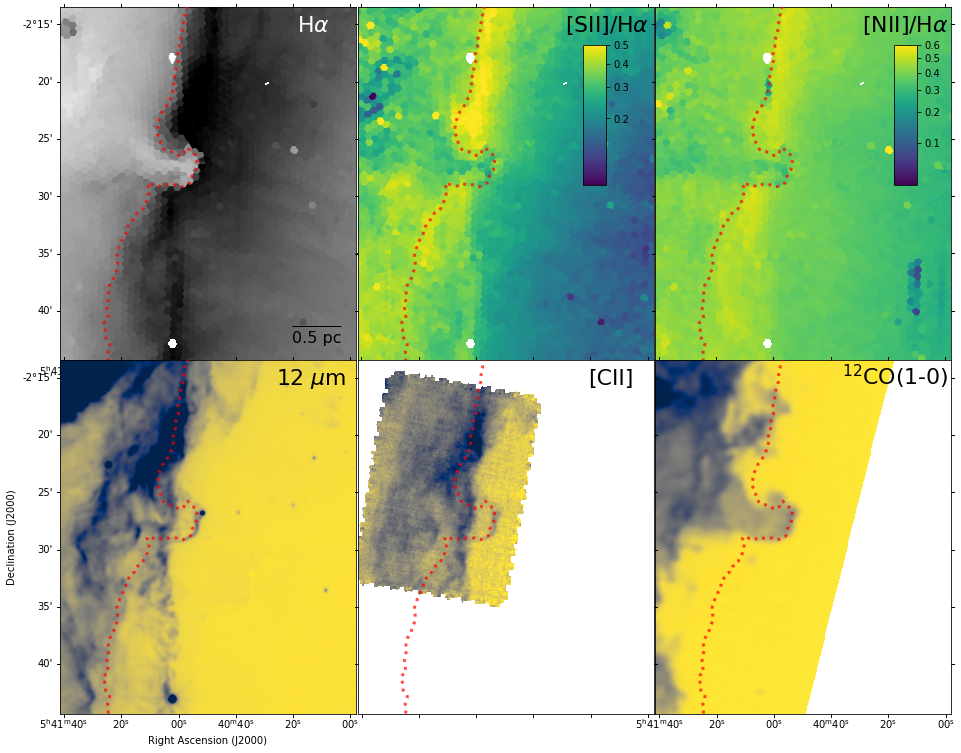}
   \caption{Comparing ionized gas with dense gas and dust in the Horsehead Nebula.   In \ha\ (top left), we see the filamentary photo-evaporation flow that streams out from the PDR towards the west.  The Horsehead sits in projection against the vertically oriented PDR, where we see increased \sii/\ha\ and \nii/\ha\ (top center and left) in the ionization front.   
   \ha\ contours (red) mirror the dust features unveiled in PAH-rich WISE 12$\mu$m imaging (bottom left),    atomic/ionized gas highlighted by \cii\ imaging (bottom center, \citealt{Pabst2017}) and molecular gas seen in $^{12}$CO(1-0) (bottom right; \citealt{Pety2017}).  
   }
              \label{fig:horsehead}%
    \end{figure*}

These prominent ionized nebulae are adjacent to the Orion~B molecular cloud (L1630), which is located directly to the east of IC~434. At the interface between the \hii\ region and the dense molecular cloud, the UV radiation from the young massive stars sustaining the ionized nebulae impinge upon the molecular gas \citep{Hollenbach1999}. This marks boundaries between ionized, atomic, and molecular gas phases, resulting in a pronounced ionization front and PDR. While this ionization front is seen prominently as a linear filament, bright in \sii/\ha\ and \nii/\ha\ (Figure \ref{fig:neb_conditions}), that extends from north to south at the eastern boundary of IC~434, pillars and other features reflecting individual inhomogeneities in the original dense material are also apparent. 

The Horsehead nebula (Barnard 33) represents one of the best known examples of these pillar-type aggregations, and has been extensively studied in this context of understanding how stellar feedback impacts the physics of molecular gas and dust in these transition regions \citep{Pety2005, Habart2005}, and the general role of UV photons in setting the chemistry and temperature structures in the ISM. 

In Figure \ref{fig:horsehead} we show how sharply the \ha\ emission traces the edge of the ionization front, with the Horsehead itself appearing as a low intensity inclusion into the IC~434 nebula. The contours delineated by our \ha\ map match with high precision the sharp edges visible in the WISE 12$\mu$m PAH sensitive bands, along with the atomic and ionized gas at the surface of the PDR traced by \cii\ 158~$\mu$m emission \citep{Pabst2017} and the dense molecular gas that remains embedded in $^{12}$CO(1-0) imaging \citep{Pety2017}. The Horsehead is seen in projection against the ionization front, which is marked by a strong increase in \sii/\ha\ and \nii/\ha\ line ratios, that sit nearer to the western ionizing source than the PAH-rich PDR and subsequent remaining dense molecular material. While a less striking example of this geometry than the better-studied Orion Bar, this region experiences a more modest UV radiation field that may be more representative of Galactic and extragalactic star-forming regions  \citep{Hernandez-Vera2023}.

The edge of the molecular cloud also marks the location of a photo-evaporative flow \citep{Henney2005}, where the radiation from $\sigma$~Ori ionizes material at the edge of the cloud, which can then accelerate and flow into the low-density bubble via a champagne flow \citep{Tenorio-Tagle1979}. This is seen in the wispy \ha\ streaks oriented perpendicular to the PDR at the plane parallel surface near the Horsehead Nebula (Figure \ref{fig:horsehead}, top left). As these motions, seen in projection, are largely in the plane of the sky we do not see any clear kinematic evidence for this flow in our \ha\ line kinematics within our instrumental limits of 5~km~s$^{-1}$ (in projection).

\begin{figure}
   \centering
   \includegraphics[width=3in]{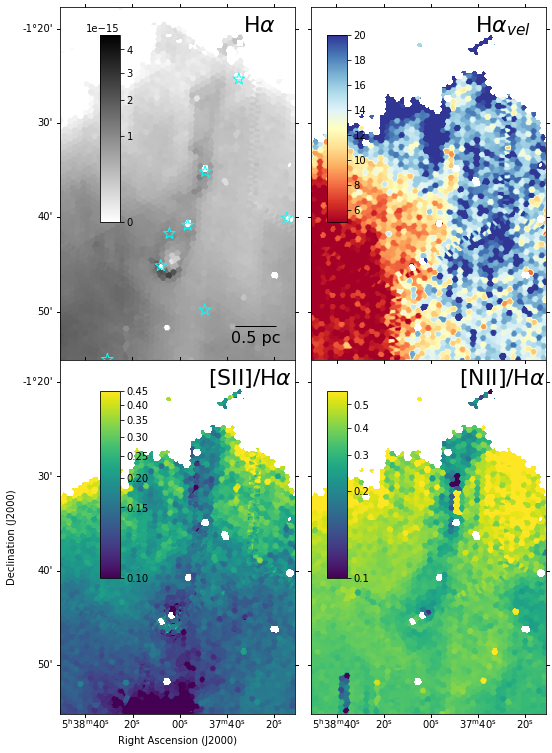}
   \caption{Ori I-2, a cometary globule to the north of IC~434. \ha\ emission (top left) traces out a $\sim$4~pc long tail, which also exhibits  lower line ratios in \sii/\ha\ (bottom left) and \nii/\ha\ (bottom right). Tentative kinematic differences are also seen in \ha\ emission tracing the tail (top right).  An overdensity of B stars (cyan) is also associated with the cometary tail. Note that the vertical stripe in the \nii/\ha\ at the center of the tail is a data artifact.}
              \label{fig:globule}%
    \end{figure}

A final set of objects visible in our ionized gas map are individual cold clumps and cometary globules \citep{PlanckXXVIII, Konyves2020}. A few are prominently seen at the edges of IC~434 (Figure \ref{fig:orion_3color}), with tails approximately oriented radially away from the central ionizing star $\sigma$~Ori. These structures resemble comet-like formations similar to the photoevaporating protoplanetary discs (proplyds) observed in various ionized nebulae \citep{ODell:1993,Mesa-Delgado:2016,MendezDelgado2022b}. We zoom in onto one of those objects, Ori I-2 \citep{Mookerjea2009}, just to demonstrate variations in optical emission line properties that can be seen in some of these objects (Figure \ref{fig:globule}). Line ratios show systematically lower values, and tentative kinematic differences are also seen in \ha\ emission tracing the $\sim$4~pc long tail, with $\sim$10~km~s$^{-1}$ lower velocities compared to the surrounding gas. This source has been proposed as an example of triggered star formation, and the tail is clearly associated with an overdensity of B-type stars, and additional YSOs are identified in the dense globule \citep{Alcala2008}.

\section{Conclusions and outlook}
\label{sec:conclusion}

The Local Volume Mapper (LVM) has begun survey operations, providing a first glimpse at the rich variety of science that will be possible by providing optical spectroscopic maps across well-studied Milky Way \hii\ regions. By providing multiple line diagnostics, it is possible to characterize the physical conditions in the gas and relate these gas-phase diagnostics to the stellar parameters of individual ionizing sources and the surrounding dense molecular gas. This paper shows first results for a 30~pc diameter region in Orion (2~deg radius), centered on three ionized nebulae (the Flame Nebula, IC~434, and IC~432). This represents $\sim$12\% of the planned contiguous coverage across the nebula (Figure \ref{fig:constellation}), based on a data reduction pipeline that is under active development and continually improving. Our conclusions can be summarized as follows:

   \begin{itemize}
      \item The \ha\ morphology is complementary to archival multi-wavelength data,  showing channels that are sculpted by ionizing photons, allowing radiation to escape into the surrounding diffuse medium. 
      \item Maps combining different line ratios are able to directly show the ionization structure of the nebulae. These can be used to infer changes in gas density and the degree of ionization. In diagnostic BPT diagrams, they can be used to track the balance of photoionization, photo-disassociation, and shocks in the ISM. 
      \item Building from the resolved to an integrated view of these nebulae, we are providing tools to connect with extragalactic studies of star-forming regions. 
      \item We highlight the potential for exciting synergies, combining ionized gas diagnostics with individual stars. Notably, in the case of IC~432, we can directly show that in addition to a reflection nebula, there is also photoionized emission associated with the central B2 spectral type star. 
      \item Multi-wavelength views link dust and dense gas with our new LVM perspectives on the ionized ISM, charting the physical conditions at the interfaces with molecular clouds. 
   \end{itemize}

These $\sim$100 LVM pointings represent less than 1\% of what LVM will achieve in its survey of the Milky Way, and only one of the hundreds of \hii\ regions that the LVM will map at sub-pc resolution, providing an inspiring view towards the rich panoptic science goals of SDSS-V.

\begin{acknowledgements}
\revone{We thank the referee for their careful reading of this work and thoughtful suggestions. } 
Funding for the Sloan Digital Sky Survey V has been provided by the Alfred P. Sloan Foundation, the Heising-Simons Foundation, the National Science Foundation, and the Participating Institutions. SDSS acknowledges support and resources from the Center for High-Performance Computing at the University of Utah. SDSS telescopes are located at Apache Point Observatory, funded by the Astrophysical Research Consortium and operated by New Mexico State University, and at Las Campanas Observatory, operated by the Carnegie Institution for Science. The SDSS web site is \url{www.sdss.org}.

SDSS is managed by the Astrophysical Research Consortium for the Participating Institutions of the SDSS Collaboration, including Caltech, The Carnegie Institution for Science, Chilean National Time Allocation Committee (CNTAC) ratified researchers, The Flatiron Institute, the Gotham Participation Group, Harvard University, Heidelberg University, The Johns Hopkins University, L’Ecole polytechnique f\'{e}d\'{e}rale de Lausanne (EPFL), Leibniz-Institut f\"{u}r Astrophysik Potsdam (AIP), Max-Planck-Institut f\"{u}r Astronomie (MPIA Heidelberg), Max-Planck-Institut f\"{u}r Extraterrestrische Physik (MPE), Nanjing University, National Astronomical Observatories of China (NAOC), New Mexico State University, The Ohio State University, Pennsylvania State University, Smithsonian Astrophysical Observatory, Space Telescope Science Institute (STScI), the Stellar Astrophysics Participation Group, Universidad Nacional Aut\'{o}noma de M\'{e}xico, University of Arizona, University of Colorado Boulder, University of Illinois at Urbana-Champaign, University of Toronto, University of Utah, University of Virginia, Yale University, and Yunnan University.

KK, OE, EE, JEMD, JL, and NS gratefully acknowledge funding from the Deutsche Forschungsgemeinschaft (DFG, German Research Foundation) in the form of an Emmy Noether Research Group (grant number KR4598/2-1, PI Kreckel) and the European Research Council’s starting grant ERC StG-101077573 (“ISM-METALS"). 

J.G.F-T gratefully acknowledges the grants support provided by Proyecto Fondecyt Iniciaci\'on No. 11220340, Proyecto Fondecyt Postdoc No. 3230001 (Sponsoring researcher), from the Joint Committee ESO-Government of Chile under the agreement 2021 ORP 023/2021 and 2023 ORP 062/2023.

C. R-Z. acknowledges support from project UNAM-PAPIIT IG101723

G.A.B. acknowledges the support from the ANID Basal project FB210003. 

AS gratefully acknowledges support by the Fondecyt Regular (project code 1220610), and ANID BASAL project FB210003.

This research has made use of the SIMBAD database,
operated at CDS, Strasbourg, France \citep{Wenger2000}.  

Multi-color images generated using the \textsc{python} package \textsc{multicolorfits} \citep{Cigan2019}. 

This publication makes use of data products from the Wide-field Infrared Survey Explorer, which is a joint project of the University of California, Los Angeles, and the Jet Propulsion Laboratory/California Institute of Technology, funded by the National Aeronautics and Space Administration.

Based in part on observations made with the NASA/DLR Stratospheric Observatory for Infrared Astronomy (SOFIA). 

The Digitized Sky Surveys were produced at the Space Telescope Science Institute under U.S. Government grant NAG W-2166. The images of these surveys are based on photographic data obtained using the Oschin Schmidt Telescope on Palomar Mountain and the UK Schmidt Telescope. The plates were processed into the present compressed digital form with the permission of these institutions.

\end{acknowledgements}

%
%

\bibliographystyle{aa}

\end{document}